\documentclass[MSNbibl,nameyear,seceqn,dvips]{arxstspdf}
\usepackage{flushend}
\usepackage{stfloats}
\usepackage{dcolumn}

\volume{29}
\issue{3}
\pubyear{2014}
\firstpage{380}
\lastpage{396}
\doi{10.1214/13-STS461} 

\makeatletter
\newcolumntype{d}[1]{D{.}{.}{#1}}
\newcommand{\convp}{\stackrel{p}{\rightarrow}}
\newtheorem{lemma}{Lemma}
\newtheorem{theorem}[lemma]{Theorem}
\newtheorem{corollary}[lemma]{Corollary}
\makeatother

\begin{document}
\begin{frontmatter}

\title{Oracle, Multiple Robust and Multipurpose Calibration in a
Missing Response Problem}
\runtitle{Multiple Modeling in Missing Data}

\begin{aug}
\author[a]{\fnms{Kwun Chuen Gary} \snm{Chan}\corref{}\ead[label=e1]{kcgchan@uw.edu}}
\and
\author[b]{\fnms{Sheung Chi Phillip} \snm{Yam}\ead[label=e2]{scpyam@sta.cuhk.edu.hk}}
\runauthor{K. C. G. Chan and S. C. P. Yam}

\affiliation{University of Washington and Chinese University of Hong Kong}

\address[a]{Kwun Chuen Gary Chan is Associate Professor,
Department of Biostatistics and Department of Health Services,
University of Washington,
Seattle, Washington 98195, USA \printead{e1}.}
\address[b]{\mbox{Sheung~Chi~Phillip~Yam} is Associate Professor,
Department of Statistics,
Chinese University of Hong Kong, Hong Kong, China \printead{e2}.}

\end{aug}

%
\begin{abstract}
In the presence of a missing response, reweighting the complete case
subsample by the inverse of nonmissing probability is both intuitive
and easy to implement. When the population totals of some auxiliary
variables are known and when the inclusion probabilities are known by
design, survey statisticians have developed calibration methods for
improving efficiencies of the inverse probability weighting estimators
and the methods can be applied to missing data analysis. Model-based
calibration has been proposed in the survey sampling literature, where
multidimensional auxiliary variables are first summarized into a
predictor function from a working regression model. Usually, one
working model is being proposed for each parameter of interest and
results in different sets of calibration weights for estimating
different parameters. This paper considers calibration using multiple
working regression models for estimating a single or multiple
parameters. Contrary to a common belief that overfitting hurts
efficiency, we present three rather unexpected results. First, when the
missing probability is correctly specified and multiple working
regression models for the conditional mean are posited, calibration
enjoys an oracle property: the same semiparametric efficiency bound is
attained as if the true outcome model is known in advance. Second, when
the missing data mechanism is misspecified, calibration can still be a
consistent estimator when any one of the outcome regression models is
correctly specified. Third, a common set of calibration weights can be
used to improve efficiency in estimating multiple parameters of
interest and can simultaneously attain semiparametric efficiency bounds
for all parameters of interest. We provide connections of a wide class
of calibration estimators, constructed based on generalized empirical
likelihood, to many existing estimators in biostatistics, econometrics
and survey sampling and perform simulation studies to show that the
finite sample properties of calibration estimators conform well with
the theoretical results being studied.
\end{abstract}

%

\end{frontmatter}\newpage

\section{Introduction}

Inverse probability weighting (IPW) was original\-ly proposed by
\citet{horvitz1952generalization} for reweighting a probability sample obtained from
a complex survey design in order to properly represent an underlying
study population. The estimator has also been widely used for missing
data problems, where complete-case data are reweighted by the inverse
of nonmissing probabilities. While inverse probability weighted
estimation is intuitive and easy to implement, the estimator is not
efficient in general and is not robust against misspecification of a
missing probability model.

In survey sampling, population totals of certain auxiliary variables
can be accurately ascertained from census data. Calibration was
proposed by Deville and S\"arndal (\citeyear{deville1992calibration}) in survey sampling literature
to utilize information from such auxiliary data. In missing data
problems, we often have a data structure similar to survey sampling
with auxiliary information. In addition to the variable of main
interest which is subject to missingness, certain covariates are
collected in the full sample to describe the missingness mechanism.
Calibration can be performed to match the moments of auxiliary
variables from the complete-case subsample to the full sample.
Nonetheless, an important difference is that calibration was originally
proposed when inclusion probability is known by design, whereas in
missing data applications the nonmissing probability is usually not
known but is being modeled and estimated from the data. In this paper,
we consider missing data problems in a sample from an infinite
population. Recently, survey calibration has been applied to study
other statistical problems; see Breslow et al. (\citeyear{breslow2009improved}),
Lumley, Shaw and Dai (\citeyear{lumley2011connections}) and \citet{saegusa2013weighted}.

When individual values of auxiliary variables are known, model
calibration can be constructed using a general working regression model
(\cite{wu2001model}). However, the methods considered in the
literature all assume a single working model for the estimation of a
single parameter. In this paper we consider multiple non-nested working
models for calibration estimation of a single or multiple parameters.
While it is a common belief that multiple modeling acts like
overfitting and the estimation efficiency should therefore be lower
compared to a single working model that is carefully chosen, we show
several surprising results that this common belief is not true for
calibration estimation. First, when the missing data probability is
correctly specified and multiple working outcome regression models are
posited, calibration enjoys an oracle property: the same semiparametric
efficiency bound is attained as if the true outcome model is known in
advance. Second, when the missingness mechanism is misspecified,
calibration can still be a consistent estimator when one of the outcome
regression models is correctly specified. Third, a common set of
calibration weights can be used to improve efficiency in estimating
multiple parameters and can simultaneously attain semiparametric
efficiency bounds for multiple parameters of interest. In fact, the
theoretical results suggest that multiple modeling can be beneficial in
practice.

The paper is organized as follows. In Section~\ref{sec2} we consider a missing
response model and define calibration estimating equations to match
moment conditions between the complete-case subsample and the full
sample. Calibration weighting is implemented using generalized
empirical likelihood (\cite{newey2004higher}) and yields weights which
are non-negative for all subjects. Sections~\ref{sec3} to~\ref{sec5} contain the main
theoretical results of this paper. In Section~\ref{sec3} we show that when the
missing data probability is correctly specified and multiple working
outcome regression models are posited, calibration enjoys an oracle
property where the same semiparametric efficiency bound is attained as
if the true outcome model is known in advance. In Section~\ref{sec4} we show
that when the missingness mechanism is misspecified, calibration can
still be a consistent estimator when one of the outcome regression
models is correctly specified. In Section~\ref{sec5} we show that a common set
of calibration weights can be used to improve efficiency in estimating
multiple parameters of interest by simultaneously calibrating to
multiple working models. Three important special cases of the
generalized empirical likelihood calibration will be discussed in
Section~\ref{sec6} and are shown to be related to many existing estimators in
the biostatistics, econometrics and survey sampling literature.
Numerical examples, including simulation studies and an analysis of
medical cost data from the Washington basic health plan, will be
presented in Section~\ref{sec7}. Discussions and several related extensions will
be presented in Section~\ref{sec8}.

\section{Calibration Estimators}\label{sec2}

In this section we consider a general framework for modifying inverse
probability weights by calibration to include information from all
observations. We consider the following missing response problem. Let
$Y$ be a random variable and $X$ be a random vector. Suppose the full
data $(y_1,x_1),\ldots,(y_N,x_N)$ are \emph{i.i.d.} from an
unspecified distribution $F_0(y,x)$. Let $R$ be a random variable
corresponding to the nonmissing indicator. The observed data can be
represented as $(r_i, r_iy_i,x_i)$, $i=1,\ldots,N$. We are interested
in estimating $\mu=E(Y)$, where $Y$ is subject to missingness and
auxiliary variables $X$ are completely observed.

We consider the case under missing at random, \emph{that is},
$P(R=1|Y,X)=P(R=1|X)=\pi_0(X)$. Suppose $P(R=1|X)=\pi(X;\beta_0)$,
where $\beta_0$ is a finite dimensional parameter. A conventional
choice of a missing data model is a logistic regression model with
linear predictors in $X$, though this is not necessary. Based on
$(r_1,x_1),\ldots,(r_N,x_N)$, the parameter $\beta_0$ can be
estimated by solving a likelihood score equation $N^{-1}\sum_{i=1}^N
s(x_i;{\beta})=0$, where $s(x;\beta)=[1-\pi(x;\beta)]^{-1}[r_i-\pi
(x;\beta)]\frac{\partial\pi}{\partial\beta}(x;\beta)$ and we
denote $\hat{\beta}$ to be the solution. When the missing data
mechanism is correctly modeled, the inverse probability weighted estimator
%
\begin{equation}
\label{E:ivee} \hat{\mu}_{\mathrm{IPW}}=\frac{1}{N}\sum
_{i=1}^N{\frac{r_i}{\pi(x_i;\hat
{\beta})}}y_i
\end{equation}
is a consistent estimator of $\mu$. However, (\ref{E:ivee}) is
generally not fully efficient because information from $\{x_i, i\dvtx
r_i=0\}$ is not utilized except in the estimation of $\beta_0$ and
such information may not be highly relevant to the estimation of $\mu
$. To improve efficiencies, we note that for an arbitrary vector
$u(x)=(u_1(x),\ldots,u_q(x))^T$ such that $E(u^T(X)u(X))$ is finite
and $E(u(X)u^T(X))$ is invertible, the two estimators $\tilde
{u}=N^{-1}\sum_{i=1}^N r_i \pi^{-1}(x_i;\hat{\beta})u(x_i)$ and
$\bar{u}=N^{-1}\sum_{i=1}^Nu(x_i)$ are both consistently estimating
the same vector, $E(u(X))$, while the latter is more efficient because
information from all observations are utilized. Instead of using
inverse probability weights in computing $\tilde{u}$ and in
(\ref{E:ivee}), we wish to find calibration weights $\{p_i, i\dvtx r_i=1\}$ such
that the following moment conditions are satisfied:
%
\begin{equation}
\label{E:cal} \bar{u}=\sum_{i=1}^N
r_i p_i u(x_i).
\end{equation}
The dimension of $u(\cdot)$ is assumed fixed and is much less than
$N$. While $u(x)$ is assumed arbitrary in the construction of the
estimator, we will discuss a choice of $u(x)$ that is optimal in
Section~\ref{sec3}. For weights satisfying (\ref{E:cal}), the calibration
weighted complete case estimate for $E(u(X))$, which is equivalent to
$\bar{u}$ by definition, is more efficient than the inverse
probability weighted estimate $\tilde{u}$ because information from all
observations is included. When $Y$ and $u(X)$ are reasonably
correlated, it is intuitive to expect that the calibration estimator
$\hat{\mu}_{\mathrm{CAL}}=\sum_{i=1}^N r_ip_iy_i$ is possibly more efficient
than the inverse probability weighted estimator (\ref{E:ivee}). The
implied weights from moment restrictions (\ref{E:cal}) can be
explicitly defined using generalized empirical likelihood (GEL)
proposed by \citet{newey2004higher}, a method originally proposed for
efficient estimation of overidentified systems of estimating equations
commonly encountered in econometrics applications. Calibration weights
proposed by Deville and S\"arndal (\citeyear{deville1992calibration})
also satisfy (\ref{E:cal})
but the method to obtain the weights was different.


The construction of the generalized empirical likelihood calibration
weights is as follows. Let $\rho(v)$ be a concave and thrice
differentiable function on $\mathbb{R}$ such that $\rho^{(1)}\neq0$,
where $\rho^{(j)}(v)=\partial^j\rho(v)/\partial v^j$ and $\rho
^{(j)}=\rho^{(j)}(0)$. As suggested by \citet{newey2004higher}, we can
replace an arbitrary $\rho(v)$ by a normalized version $-\rho
^{(2)}/(\rho^{(1)})^2\rho([\rho^{(1)}/\rho^{(2)}]v)$ such that
$\rho^{(1)}=\rho^{(2)}=-1$. This normalization will not affect the
results. The calibration weights are defined as
%
\begin{equation}
\label{E:cal1iv}\hspace*{22pt} p_i=\frac{\pi^{-1}(x_i;\hat{\beta})\rho^{(1)}(\hat{\lambda
}^T(u(x_i)-\bar{u}))}{\sum_{j=1}^N r_j\pi^{-1}(x_j;\hat{\beta
})\rho^{(1)}(\hat{\lambda}^T(u(x_j)-\bar{u}))},
\end{equation}
where
%
\begin{equation}
\label{E:cal2iv}\hspace*{22pt} \hat{\lambda}=\arg\max_{\lambda}\sum
_{i=1}^N r_i{\pi ^{-1}(x_i;
\hat{\beta})}\rho\bigl(\lambda^T\bigl(u(x_i)-\bar{u}
\bigr)\bigr).
\end{equation}
We define a calibration (CAL) estimator to be $\hat{\mu}_{\mathrm{CAL}}=\sum_{i=1}^N r_ip_iy_i$. Although $p_i$ can be defined for $i=1,\ldots,N$,
to compute the calibration estimator and its standard error, $p_i$
needs to be computed only for the subjects with $r_i=1$. By definition,
$\sum_{i=1}^N r_ip_i=1$. The moment restrictions (\ref{E:cal}) are
satisfied following the first order condition of the maximization
problem in (\ref{E:cal2iv}).

The function $\rho(\cdot)$ can be chosen from a wide class of concave
functions, and the main results in subsequent sections state that the
choice of the function $\rho(\cdot)$ does not affect consistency,
asymptotic efficiency and other properties. This is further supported
by the simulation studies in Section~\ref{sec7}. Therefore, the choice of $\rho
(\cdot)$ is a relatively minor issue. After presenting the results for
a general $\rho(v)$ in Sections~\ref{sec3}--\ref{sec5},
we extensively discuss the
following three special cases of the generalized empirical likelihood
family in Section~\ref{sec6}:
\begin{enumerate}[3.]
\item[1.]$\rho(v)=-(v-1)^2/2$.
\item[2.]$\rho(v)=\log(1-v)$.
\item[3.]$\rho(v)=-\exp(v)$.
\end{enumerate}

They are popular due to the fact that they are closely related to the
generalized method of moments (\cite*{hansen1982large};
Hansen, Heaton and Yaron,
\citeyear{hansen1996finite}),
empirical likelihood (\cite{owen1988empirical}; \cite{qin1994empirical}) and
exponential tilting (\cite*{kitamura1997information};
Imbens, Spady and Johnson,
\citeyear{imbens1998information}).
Simulations in Section~\ref{sec7} show that the three popular $\rho$ functions
give very similar results. The idea that inverse \mbox{probability} weighting
can be improved is not due to a particular choice of the $\rho$
function but to the calibration equation (\ref{E:cal}) which matches the
incomplete subsample to the complete sample. The introduction of $\rho
(\cdot)$ is needed because the calibration equation (\ref{E:cal}) is an
over-identified system of estimating equations and, therefore, the
theory of generalized empirical likelihood can be used.

In general, the calibration weights $p_i$ are not guaranteed to be
non-negative if $\lambda$ is maximized globally in (\ref{E:cal2iv}),
except in the cases where $\rho^{(1)}(v)<0$ for all $v \in\mathbb
{R}$, such as $\rho(v)=-\exp(v)$. A way to produce non-negative
weights for the whole generalized empirical likelihood family, as
suggested by \citet{newey2004higher}, is to define $\hat{\lambda}$ to
maximize the objective function in a restricted set $\Lambda=\{\lambda
\in\mathbb{R}^q\dvtx \lambda^T(u_i(x_i)-\bar{u})\in\mathcal{V}, i\dvtx
r_i=1\}$, where $\mathcal{V}\subset\mathbb{R}$ is an open interval
containing zero. When we choose $\mathcal{V}$ to be a sufficiently
small neighborhood around zero, $p_i$ will be non-negative for all
complete-case observations. When the missing data model is correctly
specified, it follows from \citet{newey2004higher} that the restricted
maximum exists with probability approaching 1 when $N$ is large and is
asymptotically equivalent to the unrestricted maximizer. The restricted
maximization is implemented in the \texttt{gmm} package in R
(Chauss\'e, \citeyear{chausse'2010computing}).

In econometrics, generalized empirical likelihood is often employed for
estimating a $p$-dimensional parameter by specifying a $q$-dimensional
estimating \mbox{equation}, where $q>p\geq1$. However, we are not estimating
the target parameter $\mu$ by directly solving an overidentified
estimating equation. In fact, we use the moment conditions (\ref{E:cal}) to
generate weights $p_i$, which are implied weights from the generalized
empirical likelihood (\cite{newey2004higher}). The calibration
conditions (\ref{E:cal}) can be regarded as a $q$-dimensional moment
restriction with a degenerate parameter, and (\ref{E:cal2iv}) is essentially a
degenerate case of generalized empirical likelihood with only the
auxiliary parameters $\lambda$ appearing but not the target
parameters. Even though the generalized empirical likelihood estimation
problem is undefined because the moment restrictions are not functions
of target parameters, implied weights can still be constructed by
(\ref{E:cal1iv}). In econometrics, the generalized empirical likelihood estimators
are usually solutions to saddlepoint problems\vspace*{1pt} and can be difficult to
compute. In our case, $\hat{\lambda}$ is a solution to a convex
maximization problem rather than a saddlepoint problem and can be
computed by a fast and stable algorithm.


\section{Oracle Property}\label{sec3}

In Sections~\ref{sec3}--\ref{sec5} we will examine statistical properties of calibration
estimators in the context of missing data analysis. In this section we
show that the class of estimators enjoy an oracle property. We consider
model-based calibration where the functions $u(x)$ in the moment
condition (\ref{E:cal}) may depend on a finite dimensional parameter.
Let $u_1(X;\gamma_1), \ldots, u_q(X;\gamma_q)$ be $q$ non-nested
working outcome regression\vspace*{1pt} models for $E(Y|X)$ and $\gamma_0=(\gamma
_1^T,\ldots,\gamma_q^T)^T$. The parameters $\gamma_k \in\mathbb
{R}^{p_k},k=1,\ldots,q$ can be of different dimensions, and $\gamma
_0\in\mathbb{R}^p$, where $p=p_1+\cdots+p_q$. Let $\hat{\gamma
}=(\hat{\gamma}_1^T,\ldots,\hat{\gamma}_q^T)^T$ be an estimate of
$\gamma_0$. For example, $\hat{\gamma}_r$ can be a least squares
estimate for the $r${th} working model for $E(Y|X)$, $r=1,\ldots,q$.
We denote the sample mean estimate $\bar{u}(\hat{\gamma})=N^{-1}\sum_{i=1}^N u(x_i;\hat{\gamma})$ and the calibration weights satisfy
$\bar{u}(\hat{\gamma})=\sum_{i=1}^N r_ip_iu(x_i;\hat{\gamma})$,
which are found by (\ref{E:cal1iv}) and (\ref{E:cal2iv}) with $u(x)$
and $\bar{u}$ replaced by $u(x;\hat{\gamma})$ and $\bar{u}(\hat
{\gamma})$ respectively. Let $m(X;\gamma_0)=c_0+\sum_{j=1}^qc_ju_j
(X;\gamma_j)$, where $c_0,\ldots,c_q$ minimizes\vspace*{-1pt}
%
\begin{equation}
\label{E:firstorder} E \Biggl(\Biggl(Y-c_0-\sum
_{j=1}^qc_j u_j(X;
\gamma_j)\Biggr)^2 \Biggr).
\end{equation}
That is, $m(X;\gamma_0)$ is the best linear predictor of $Y$ by
$u(X;\gamma_0)$. Supposing the missing data model is correctly
specified, that is, $\pi_0(X)=\pi(X;\beta_0)$, we have the following lemma:

\begin{lemma}\label{lem1}
Under the regularity conditions stated in the supplemental article
(Chan and Yam, \citeyear{chan2013supplement}),
%
\begin{eqnarray}
\label{E:misscor} \hat{\mu}_{\mathrm{CAL}}-\mu&=&\frac{1}{N}\sum
_{i=1}^N \biggl[\frac{r_i}{\pi
_0(x_i)}
\bigl(y_i-\tilde{m}(x_i;\gamma_0)\bigr)
\nonumber
\\
&&\phantom{\hspace*{52pt}}{}+
\bigl(\tilde{m}(x_i;\gamma_0)-\mu \bigr)
\biggr]
\\
&&{}+o_p\bigl(N^{-1/2}\bigr),\nonumber
\end{eqnarray}
where
\begin{eqnarray*}
\tilde{m}(X;\gamma_0)&=&m(X;\gamma_0)
\\
&&{}-A_2^TS^{-1}
\bigl(1-\pi _0(X)\bigr)^{-1}\frac{\partial\pi}{\partial\beta}(X;
\beta_0),
\\
A_2&=&-E \biggl(\frac{\partial\pi}{\partial\beta}(X;\beta_0)
\frac
{1}{\pi(X;\beta_0)}\bigl(Y-m(X)\bigr) \biggr)
\end{eqnarray*}
and
\begin{eqnarray*}
S&=&E \biggl(\pi_0^{-1}(X) \bigl(1-\pi_0(X)
\bigr)^{-1}
\\
&&\phantom{E \biggl(}{}\cdot\frac{\partial\pi
}{\partial\beta}(X;\beta_0)\frac{\partial\pi}{\partial\beta
}^T(X;\beta_0) \biggr).
\end{eqnarray*}
\end{lemma}

A detailed proof of the lemma is given in the supplemental article
(\cite{chan2013supplement}). The above lemma holds for arbitrary sets of
functions $u(\cdot)$ satisfying mild regularity conditions. The
asymptotic representation given in Lemma~\ref{lem1} also suggests the following
plugged-in estimator for asymptotic variance:
\[
\frac{1}{N^2}\sum_{i=1}^N \biggl[
\frac{r_i}{\pi(x_i;\hat{\beta
})}\bigl(y_i-\hat{\tilde{m}}(x_i)\bigr)+
\bigl(\hat{\tilde{m}}(x_i)-\hat{\mu }_{\mathrm{CAL}}\bigr)
\biggr]^2,
\]
where
\begin{eqnarray*}
\hat{\tilde{m}}(X)&=&m(X;\hat{\gamma})
\\
&&{}-\hat{A}_2^T\hat
{S}^{-1}\bigl(1-\pi(X;\hat{\beta})\bigr)^{-1}
\frac{\partial\pi}{\partial
\beta}(X;\hat{\beta}),
\\
\hat{A}_2&=&\frac{1}{N}\times\sum
_{i=1}^N \frac{r_i}{\pi(X;\hat
{\beta})^2}\frac{\partial\pi}{\partial\beta}(x_i;
\hat{\beta }) \bigl(y_i-m(x_i;\hat{\gamma})\bigr)
\end{eqnarray*}\eject\noindent
and
\begin{eqnarray*}
\hat{S}&=&\frac{1}{N}\times\sum_{i=1}^N
\pi^{-1}(x_i;\hat{\beta }) \bigl(1-\pi(x_i;
\hat{\beta})\bigr)^{-1}
\\
&&\phantom{\frac{1}{N}\times\sum_{i=1}^N}{}\cdot\frac{\partial\pi}{\partial\beta
}(x_i;\hat{\beta})
\frac{\partial\pi}{\partial\beta}^T(x_i;\hat {\beta}).
\end{eqnarray*}

The asymptotic expansion (\ref{E:misscor}) depends on the choice of
$u(X;\gamma_0)$ implicitly through $m(X;\gamma_0)$ and we may choose
a particular $u(X;\gamma_0)$ to minimize the asymptotic variance. Let
$m_0(X)$ denote the true conditional expectation $E(Y|X)$. The
optimality properties are stated in the following theorem.

\begin{theorem}[(Semiparametric efficiency)]\label{the2}
Suppose that the regularity conditions in Lemma~\ref{lem1} hold and suppose
there exist $a_0,\ldots,a_q$ such that
%
\begin{equation}
\label{E:drcond} m_0(X)=a_0+\sum
_{j=1}^q a_j u_j(X;
\gamma_0).
\end{equation}
Then, $\sqrt{N}(\hat{\mu}_{\mathrm{CAL}}-\mu)$ converges in distribution to
$N(0,V_{\mathrm{semi}})$, where $V_{\mathrm{semi}}$ attains the semiparametric variance
bound as in Robins and
Rotnitzky (\citeyear{robins1995semiparametric}) and Hahn
(\citeyear{hahn1998role}),
\[
V_{\mathrm{semi}}=\operatorname{Var} \biggl[\frac{RY}{\pi_0(X)}- \biggl(
\frac{R}{\pi
_0(X)}-1 \biggr)m_0(X)-\mu \biggr].
\]
\end{theorem}



The proof of the theorem is given in the supplementary article (\cite{chan2013supplement}).
In Theorem~\ref{the2} the constants $a_0,\ldots,a_q$ are
arbitrary and do not need to be estimated. Theorem~\ref{the2} states that
semiparametric efficiency is attained under a condition weaker than
requiring the calibration function $u(X)$ to be identical to the true
conditional expectation $m_0(X)$; see Section~2.3 of
\citet{qin2007empirical} for a related discussion. Also,
as suggested by \citet{qin2007empirical}, we can plot $Y$
against each component of $X$ to suggest a
functional form for $u(X)$. An important implication of the theorem, an
oracle property, is given as follows. Suppose $u_1(X;\gamma_1),\ldots
,u_q(X;\gamma_q)$ are $q$ working models for $E(Y|X)$ and that one of
them, without loss of generality, say, $u_1(X;\gamma_1)$, is the true
conditional expectation.

\begin{corollary}[(Oracle property)]\label{cor3}
Under conditions in Lemma~\ref{lem1}, suppose $E(Y|X)=u_1(X;\gamma_1)$. The
estimator $\hat{\mu}_{\mathrm{CAL},1}$ where $u=u_1$ achieves the same
semiparametric efficiency bound as the estimator $\hat{\mu}_{\mathrm{CAL},2}$
where $u=(u_1,\ldots,u_q)$.
\end{corollary}

While overfitting should be avoided in usual statistical practice, and
assuming multiple working regression models have a similar flavor to
overfitting, the oracle property states that the asymptotic efficiency
of calibration estimators is not affected by multiple working models
and attains the same semiparametric efficiency bound as if the true
model is known in advance. Note that overfitting is problematic for the
estimation of regression coefficients, and we are interested in
estimating the mean of $Y$, which is a different estimand. Therefore,
the oracle property does not contradict existing statistical theory. In
Section~\ref{sec7} we show in simulation studies that multiple modeling loses a
negligible amount of efficiency even for practical sample sizes.

We would like to remark that there are substantial differences between
the oracle property for calibration estimators and the oracle property
discussed in the model selection literature. In the model selection
literature, oracle properties are often enjoyed by regularized
estimators (see, e.g., \cite*{fan2001variable} and \cite*{zou2006adaptive}), which
add a penalization term to likelihood-type functions. The purpose of
regularization is to determine nonzero coefficients from a large number
of predictors in a regression setting, and the degree of regularization
is controlled by a tuning parameter. In those situations, oracle
properties mean that when a tuning parameter is asymptotically
increasing at a certain rate smaller than $\sqrt{N}$, the regularized
estimator for the nonzero coefficients will attain the same asymptotic
variance as if the true set of nonzero coefficients are known in
advance. This property is closely related to Hodges' superefficient
estimator (\cite{lehmann1998theory}). The main differences between the
oracle property of calibration estimators and that in the model
selection literature are given as follows. First, our methods apply to
the estimation of $\mu=E(Y)$, not to estimation of the coefficients of
$E(Y| X)$. Moreover, our methods are based on weighting observations
and not by regularization of likelihood functions. Furthermore, there
is no tuning parameter to be specified with a user-defined rate of
convergence in our method.


\section{Multiple Robustness}\label{sec4}

In this section we consider the validity of calibration estimators
under misspecified missing data models. In this case, the estimator
$\hat{\beta}$ will converge in probability to some constant vector
$\beta^*$ that minimizes the Kullback--Leibler Information Criterion
(\cite{white1982maximum}), but $\pi(X;\beta^*)\neq\pi_0(X)$. When the missing
data mechanism is misspecified, the estimate $\hat{\lambda}$ will not
converge in probability to $0$ in general, but will instead converge in
probability to $\lambda^*$, where
\[
\lambda^*=\arg\max_{\lambda} E\bigl(R\pi^{-1}\bigl(X;
\beta^*\bigr)\rho\bigl\{\lambda \bigl[u(X)-u_\mu\bigr]\bigr\}\bigr),
\]
$u_\mu=E(u(X))$. We define $\tilde{w}(x)=\pi^{-1}(x;\beta^*)\times\break \rho\{
\lambda^*[u(x)-u_\mu]\}/k$, where $k=E(R\pi^{-1}(X;\beta^*)\* \rho\{
\lambda^*[u(X)-u_\mu]\})$,
%
\begin{eqnarray*}
\hspace*{-5pt}&&f({\lambda},{\beta},{\gamma})
\\
\hspace*{-5pt}&&\quad = \frac{1}{N}\sum_{i=1}^{N}
r_i \Biggl( {\pi^{-1}(x_i,{\beta
})\rho'\bigl({\lambda}\bigl(u(x_i,{\gamma})-\bar{u}({\gamma})\bigr)\bigr)}
\\
\hspace*{-5pt}&&\phantom{\quad = \frac{1}{N}\sum_{i=1}^{N}\Biggl(\ \ }{}\cdot\Biggl( N^{-1}
\sum_{i=1}^{N} r_j\pi^{-1}(x_j,{\beta})
\\
\hspace*{-5pt}&&\phantom{\hspace*{110pt}}{}\cdot\rho'\bigl(\hat{\lambda
}\bigl(u(x_j,{\gamma})-\bar{u}({\gamma})\bigr)\bigr)\Biggr)^{-1}
\\
\hspace*{-5pt}&&\phantom{\quad = \frac{1}{N}\sum_{i=1}^{N}
r_i \biggl(\hspace*{106pt}}{}-\pi^{-1}(x_i,{
\beta}) \Biggr)
\\
\hspace*{-5pt}&&\phantom{\quad = \frac{1}{N}\sum_{i=1}^{N}}{}\cdot\bigl(y_i-m(x_i,{\gamma})\bigr)
\\
\hspace*{-5pt}&&\qquad{}+\frac{1}{N}\sum_{i=1}^{N}
\biggl[\frac{r_i}{\pi(x_i,{\beta
})}\bigl(y_i-m(x_i,{\gamma})
\bigr)
\\
\hspace*{-5pt}&&\phantom{\qquad{}+\frac{1}{N}\sum_{i=1}^{N}\hspace*{32pt}}{}+\bigl(m(x_i,{\gamma})-\mu\bigr) \biggr]
\end{eqnarray*}
%
and $f_0(\lambda,\beta,\gamma)=E(f(\lambda,\beta,\gamma))$.
%
\begin{theorem}[(Robustness)]\label{the4}
Suppose the missing data model is misspecified but
condition (\ref{E:drcond}) holds for the calibration function
$u(X;\gamma_0)$, the regularity conditions in Lemma~\ref{lem1} hold,\vspace*{2pt} and
$E[\sup_{(\lambda,\beta,\gamma)}|f(\lambda,\beta,\allowbreak  \gamma
)|]<\infty$. Then, the calibration estimator $\hat{\mu}_{\mathrm{CAL}}$ is a
consistent estimator for $\mu$.
\end{theorem}

The proof is as follows:
\begin{eqnarray*}
\hat{\mu}_{\mathrm{CAL}}&=&\sum_{i=1}^N
r_ip_i\Biggl(y_i-\Biggl(a_0+
\sum_{j=1}^q a_j
u_j(x_i;\hat{\gamma})\Biggr)\Biggr)
\\
&&{}+\sum
_{i=1}^N r_ip_i
\Biggl(a_0+\sum_{j=1}^q
a_j u_j(x_i;\hat{\gamma})\Biggr)
\\
&=&\sum_{i=1}^N r_i
p_i\Biggl(y_i-\Biggl(a_0+\sum
_{j=1}^q a_j u_j(x_i;
\hat {\gamma})\Biggr)\Biggr)
\\
&&{}+\frac{1}{N} \sum_{i=1}^N
\Biggl(a_0+\sum_{j=1}^q
a_j u_j(x_i;\hat{\gamma})\Biggr)
\\
&=&\sum_{i=1}^N {r_i\pi^{-1}(x_i;\hat{\beta})\rho\bigl(\hat
{\lambda}\bigl(u(x_i)-\bar{u}\bigr)\bigr)}
\\
&&\phantom{\sum_{i=1}^N}{}\cdot\Biggl(\sum_{j=1}^N r_j\pi^{-1}(x_j;\hat
{\beta})\rho\bigl(\hat{\lambda}\bigl(u(x_j)-\bar{u}\bigr)\bigr)\Biggr)^{-1}
\\
&&\phantom{\sum_{i=1}^N}{}\cdot
\Biggl(y_i-\Biggl(a_0+\sum_{j=1}^q
a_j u_j(x_i;\hat{\gamma})\Biggr)\Biggr)
\\
&&{}+\frac{1}{N} \sum_{i=1}^N
\Biggl(a_0+\sum_{j=1}^q
a_j u_j(x_i;\hat{\gamma})\Biggr)
\\
&\convp&E\bigl(R\tilde{w}(X) \bigl(Y-m_0(X)\bigr)\bigr)+E
\bigl(m_0(X)\bigr)
\\
&=&E\bigl(\pi_0(X)\tilde{w}(X) \bigl(E(Y|X)-m_0(X)
\bigr)\bigr)
\\
&&{}+E\bigl(E(Y|X)\bigr)
\\
&=&0+\mu=\mu.
\end{eqnarray*}

The first equality holds by adding and subtracting the same quantity,
the second equality holds because of (\ref{E:cal}), the third equality holds by
the definition of $p_i$, and the convergence in probability holds by
the convergence of plugged-in estimates and the uniform convergence of
$f(\lambda,\beta,\gamma)$ guaranteed by the regularity conditions,
and the last line holds because $E(Y|X)=m_0(X)$. An immediate corollary
is that when one of the $q$ working models for $E(Y|X)$ is correctly
specified, the calibration estimator is consistent even when the
missing data model is misspecified. Therefore, calibration estimators
enjoy the following multiple robust property: consistency holds when
either the missing data model or any one of the working outcome
regression models is correctly specified. Doubly robust estimators
(e.g., augmented inverse probability weighted estimators) have been
popular in missing data analysis because of their extra protection
against misspecification of the missing data model. However, a single
working outcome regression model may be misspecified as well. Double
robustness of calibration estimators has been discussed recently in
\citet{kott2010using}. Our results show further that calibration
estimators allow multiple non-nested working models to be assumed and
is consistent when any one of the working models are correctly
specified. This provides an even better protection against model
misspecification than the existing doubly robust estimators.

\section{Multipurpose Calibration}\label{sec5}

Very often, in addition to the sample mean, we are also interested in
estimating other functionals of the distribution of $Y$, $F_0(y)$, for
example, the proportion of units with an outcome value no more than $t$,
\[
F_0(t)=\int_{-\infty}^t
dF_0(y)=\int I(y\leq t)\,dF_0(y).
\]
For $L$ functions $h_1,\ldots,h_L\dvtx \mathbb{R}\to\mathbb{R}$, let
$\mu_l=\break   \int h_l(y)\,dF_0(y)$, $l=1,\ldots,L$ be $L$ parameters of
interest. To estimate $\mu_l$, we may posit a working model $m_l(X)$
for $E(h_l(Y)|X)$, and calibration weights $p_{li}$ can be found by
(\ref{E:cal1iv}) and (\ref{E:cal2iv}). A calibration estimator for
$\mu_l$ can then be defined as $\sum_{i=1}^N r_ip_{li} h_l(y_i)$.
However, the set of weights $\{p_{li}\}$ are different for each
estimand. When the construction of weights and the analysis are done by
different statisticians, the use of multiple sets of weights may not be
practical. Moreover, a set of weights that is optimal for estimating
one particular parameter is likely to be suboptimal for estimating
other parameters.

We would like to use the same set of weights to estimate $\mu_1,\ldots
, \mu_L$ simultaneously. To do this, we find the weights by (\ref
{E:cal1iv}) and (\ref{E:cal2iv}) with $u=(m_1,\ldots,m_L)^T$, that
is, to calibrate to the $L$ working models for different conditional
expectations simultaneously. Working models can be suggested by
exploratory data analysis, prior scientific knowledge or by convention.
For instance, if $h_l(Y)=I(Y>c)$ for some constant $c$, one may use a
logistic regression model with a linear predictor in $X$ for $m_l$. By
calibration to $u=(m_1,\ldots,m_L)^T$, we obtain a common set of
weights. The estimates for $\mu_1,\ldots,\mu_L$ are defined as
\[
\hat{\mu}_1=\sum_{i=1}^Nr_ip_i
h_1(y_i),\quad \ldots, \quad\hat{\mu}_L=
\sum_{i=1}^Nr_ip_i
h_L(y_i).
\]
We have the following theoretical properties of the estimators.
%
\begin{theorem}\label{the5}
Suppose $\pi(X;\beta)$ is correctly specified, the regularity
conditions stated in Lemma~\ref{lem1} hold, and assume that $E(h_l^2(Y))<\infty
$ for $l=1,\ldots,L$. We have the following properties:
\begin{enumerate}[(b)]
\item[(a)] The estimates $\hat{\mu}_1,\ldots,\hat{\mu}_L$ are all
consistent for $\mu_1,\ldots,\mu_L$, regardless of the validity of
working models $m_l(X)$.
\item[(b)] When $m_l(X)=E(h_l(Y)|X)$, for $1\leq l \leq j \leq L$,
$\hat{\mu}_1,\ldots,\hat{\mu}_j$ are asymptotically semiparametric
efficient.
\end{enumerate}
\end{theorem}

Statement (a) in the above theorem can be proven using similar
arguments as in Lemma~\ref{lem1} and statement (b) follows from Corollary~\ref{cor3}.
Theorem~\ref{the5} states that a common set of calibration weights can be used
to improve efficiency
in estimating multiple parameters of interest by
simultaneously calibrating to multiple working models.

In practice, the construction of weights and the estimation of target
parameters may be performed by different statisticians. The
statistician who constructs the weights may not know which estimand is
of ultimate interest. Suppose the parameter of interest is $E(h(Y))$.
Since $E(h(Y))$ is a Riemann--Stieltjes integral, we can use the
discrete approximation
\begin{eqnarray*}
&&\int h(y)\,dF_0(y)
\\
&&\quad \approx  \sum_{m=0}^M
h \biggl(\frac
{t_m+t_{m+1}}{2} \biggr)\int I(t_m< y \leq
t_{m+1})\,dF_0(y)
\\
&&\quad = \sum_{m=0}^M h \biggl(
\frac{t_m+t_{m+1}}{2} \biggr)\bigl[F_0(y_{m+1})-F_0(y_m)
\bigr]
\end{eqnarray*}
to approximate arbitrary $E(h(Y))$, where $-\infty\equiv
t_0<t_1<t_2<\cdots<t_M<\infty\equiv t_{M+1}$. The parameter of
interest, $E(h(Y))$, can therefore be approximated by a linear
combination of $[F_0(t_{i+1})-F_0(t_{i})]$. We can construct working
models for $P(t_m<Y\leq t_{m+1}| X)$ to improve the estimation of
$[F_0(t_{i+1})-F_0(t_{i})]$, and the estimation of $E(h(Y))$ can be
improved by calibrating to $M+1$ models for $P(t_m<Y\leq t_{m+1}|
X)$, $m=0,\ldots,M$.

\section{Special Cases and Relationship to Existing Estimators}\label{sec6}

In this section we consider several special cases of the generalized
empirical likelihood calibration estimator and discuss their
connections to existing estimators proposed in biostatistics,
econometrics and survey sampling.

When $\rho$ is a quadratic function, after normalization we have $\rho
^{(1)}(v)=-v-1$. From (\ref{E:cal2iv}), $\hat{\lambda}$ has an
explicit solution,
\begin{eqnarray*}
\hat{\lambda}&=&- \Biggl[\sum_{i=1}^N
r_i\pi^{-1}(x_i,\hat{\beta })
\bigl(u(x_i)-\bar{u}\bigr)^{\otimes2} \Biggr]^{-1}
\\
&&{}\cdot\Biggl[\sum_{i=1}^N r_i
\pi^{-1}(x_i,\hat{\beta}) \bigl(u(x_i)-\bar{u}
\bigr) \Biggr],
\end{eqnarray*}
where for a row vector $a$, $a^{\otimes2}=aa^T$. The calibration
estimator is equivalent to
%
\begin{eqnarray}
\label{E:aipw} \hspace*{10pt}\hat{\mu}_{\mathrm{CAL}, \mathrm{Q}}&=&\frac{\sum_{i=1}^N r_i\pi^{-1}(x_i;\hat{\beta
})[y_i-c_1^Tu(x_i)]}{\sum_{i=1}^N r_i \pi^{-1}(x_i;\hat{\beta
})}
\nonumber
\\[-8pt]
\\[-8pt]
&&{}+c_1^T
\frac{1}{N}\sum_{i=1}^N
u(x_i),\nonumber
\end{eqnarray}
where
\begin{eqnarray*}
c_1&=&\sum_{i=1}^N
r_i\pi^{-1}(x_i,\hat{\beta})
\\
&&{}\cdot\Biggl[\sum
_{i=1}^N r_i
\pi^{-1}(x_i,\hat{\beta}) \bigl(u(x_i)-\bar{u}
\bigr)^{\otimes2} \Biggr]^{-1}
\\
&&{}\cdot \bigl[ \bigl(u(x_i)-
\bar{u}\bigr)y_i \bigr].
\end{eqnarray*}
This special case of the generalized empirical likelihood calibration
estimator corresponds to the generalized regression estimator (Cassel, S{\"a}rndal and
Wretman,
\citeyear{cassel1976some}).
The quadratic generalized empirical \mbox{likelihood} is also
closely related to the quadratic likelihood discussed in
\citet{lindsay2003inference}. Note that when the missingness model is correctly specified,
the denominator $\sum_{i=1}^N r_i \pi^{-1}(x_i;\hat{\beta})$ on the
left-hand side of (\ref{E:aipw}) is approximately $N$, so the
estimator (\ref{E:aipw}) is also similar to the augmented inverse
probability weighted (AIPW) estimating equation proposed by
Robins, Rotnitzky and Zhao
(\citeyear{robins1994estimation}). Breslow et al. (\citeyear{breslow2009improved})
and
Lumley, Shaw and Dai
(\citeyear{lumley2011connections}) discussed
the connections between the augmented inverse probability weighted and
the calibration estimators. A~related regression-based doubly robust
estimator was discussed in
Scharfstein, Rotnitzky and Robins
(\citeyear{scharfstein1999adjusting}) and \citet{bang2005doubly},
and extended to a multiple robust estimator in \citet{chan2013a}.
However, these estimators were constructed from a different
framework and do not have associated calibration weights.

Empirical likelihood (EL) is another special case of the generalized
empirical likelihood which is frequently studied in the literature
(\cite{owen1988empirical}; \cite{qin1994empirical}) and which corresponds\vadjust{\goodbreak} to $\rho
(v)=\log(1-v)$. In this case, $\hat{\lambda}$ is a solution to the
system of equations
\[
\sum_{i=1}^N \frac{r_i\pi^{-1}(x_i;\hat{\beta})(u(x_i)-\bar
{u})}{1-\lambda^T(u(x_i)-\bar{u})}=0
\]
and
\[
p_i=\frac{[\pi(x_i;\hat{\beta})(1-\hat{\lambda}^T(u(x_i)-\bar
{u}))]^{-1}}{\sum_{j=1}^N r_i[\pi(x_j;\hat{\beta})(1-\hat{\lambda
}^T(u(x_j)-\bar{u}))]^{-1}}.
\]
The empirical likelihood calibration has a pseudo nonparametric maximum
likelihood interpretation, where $p_i$ maximizes a weighted
loglikelihood\break  $\sum_{i=1}^N r_i\* \pi^{-1}(x_i;\hat{\beta})\log p_i$
subject to the moment condition (\ref{E:cal}). Moment matching using
empirical likelihood has been discussed in
\citet{hellerstein1999imposing}, \citet{tan2006distributional},
\citet{qin2007empirical}, \citet{chan2012uniform},
Graham, De~Xavier~Pinto and Egel
(\citeyear{graham2012inverse}) and \citet{han2013estimation}.
\citet{han2013estimation} showed that the
empirical likelihood estimator of \citet{qin2007empirical} is multiply
robust, based on a property of $\rho(v)=\log(1-v)$ which is not
extensible to other members of the generalized empirical likelihood
family. In survey sampling, the empirical likelihood-based method has
been proposed to calibrate design-based weights to auxiliary data by
\citet{chen1999pseudo}, \citet{wu2001model},
Chen, Sitter and Wu (\citeyear{chen2002using}) and
\citet{kim2009calibration}, among others.

Exponential tilting (ET) is also a special case of generalized
empirical likelihood where $\rho(v)=-\exp(v)$
(\cite*{kitamura1997information};
Imbens, Spady and Johnson,
\citeyear{imbens1998information}). In this case, $\hat{\lambda}$ is a
solution of the system of equations
\[
\sum_{i=1}^N r_i
\pi^{-1}(x_i;\hat{\beta}) \bigl(u(x_i)-\bar{u}
\bigr)\exp \bigl(\lambda^T\bigl(u(x_i)-\bar{u}\bigr)
\bigr)=0
\]
and
\[
p_i=\frac{\pi^{-1}(x_i;\hat{\beta})\exp(\hat{\lambda
}^T(u(x_i)-\bar{u}))}{\sum_{j=1}^N r_i\pi^{-1}(x_j;\hat{\beta
})\exp(\hat{\lambda}^T(u(x_j)-\bar{u}))}.
\]
The estimator can also be formulated by maximizing a weighted entropy
function $\sum_{i=1}^N r_i\pi^{-1}(x_i;\hat{\beta})p_i\log p_i$
subject to the moment condition (\ref{E:cal}). This corresponds to the
raking estimators (\cite*{deming1940least};
Deville, S{\"a}rndal and Sautory,
\citeyear{deville1993generalized};
\cite*{hainmueller2012entropy}) in the survey sampling literature, and an advantage
of using the exponential tilting estimator is that the resulting
weights $p_i$ are always non-negative.

The class of generalized empirical likelihood calibration estimators
contains many more estimators than the three special cases mentioned
above. For example, the family of power divergence statistics of
\citet{cressie1984multinomial} is a proper subclass of the generalized
empirical likelihood, where for some scalar $\theta$,
\[
\rho(v)=-(1+\theta v)^{(\theta+1)/\theta}/(\theta+1).
\]
The empirical likelihood and exponential tilting estimators correspond
to the limits as $\theta\to-1$ and $\theta\to0$ respectively, and
the quadratic estimator corresponds to $\theta=1$. Several other cases
have also been considered in the literature, for example, $\theta
=-\frac{1}{2}$ (Freeman--Tukey), $\theta=-2$ (Neyman) and $\theta
=\frac{2}{3}$ (Cressie--Read).

\section{Numerical Studies}\label{sec7}

\subsection{Simulated Data}

In this section we present simulation studies and an analysis of the
Washington basic health plan data to study the finite sample
performance of the calibration estimators. The first simulation study
followed a scenario in \citet{kang2007demystifying}
for the estimation of
the population mean. The scenario was designed
so that the assumed outcome regression and missing data models were
nearly correct under misspecification, but the augmented inverse
probability weighted estimator can be severely biased.
Sample sizes for each simulated data set were 200 or 1000, and 1000
Monte Carlo data sets were generated. For each observation, a random
vector $Z=(Z_1,Z_2,Z_3,Z_4)$ was generated from a standard multivariate
normal distribution, and transformations $X_1=\exp(Z_1/2),
X_2=Z_2/(1+\exp(Z_1)), X_3=(Z_1Z_3/25+0.6)^3$ and $X_4=(Z_2+Z_4+20)^2$
were defined with $X=(X_1,X_2,X_3,X_4)$. The outcome of interest $Y$
was generated from a normal distribution with mean
$210+27.4Z_1+13.7Z_2+13.7Z_3+13.7Z_4$ and unit variance, and $Y$ was
observed with probability $\exp(\eta_0(Z))/(1+\exp(\eta_0(Z)))$,
where $\eta_0(Z)=-Z_1+0.5Z_2-0.25Z_3-0.1Z_4$. The correctly specified
outcome and missing data models were regression models with $Z$ as
covariates, whereas we treated $X$ to be the covariates instead of $Z$
in misspecified models. \citet{kang2007demystifying} showed that the
misspecified models were nearly correctly specified.

\begin{table*}
\tablewidth=\textwidth
\tabcolsep=0pt
\caption{Comparisons among the calibration estimators and other
estimators under the Kang and Schafer scenario, \textup{(a)} models in Z, \textup{(b)}
models in X. SSE represents the sampling standard deviation, RMSE
represents the root mean squared error and RE~represents~relative~efficiency~which~is the RMSE relative to $\hat{\mu}_{\mathrm{OLS}}$}\label{tab1}
\begin{tabular*}{\textwidth}{@{\extracolsep{\fill}}ld{2.2}d{2.2}d{2.2}cd{3.2}d{3.2}d{3.2}d{2.2}@{}}
\hline
& \multicolumn{4}{c}{\textbf{(a)}} & \multicolumn{4}{c@{}}{\textbf{(b)}}
\\[-5pt]
& \multicolumn{4}{c}{\hrulefill} & \multicolumn{4}{c@{}}{\hrulefill}
\\
& \multicolumn{1}{c}{\textbf{Bias}} & \multicolumn{1}{c}{\textbf{SSE}}
& \multicolumn{1}{c}{\textbf{RMSE}} & \multicolumn{1}{c}{\textbf{RE}}
& \multicolumn{1}{c}{\textbf{Bias}} & \multicolumn{1}{c}{\textbf{SSE}}
& \multicolumn{1}{c}{\textbf{RMSE}} & \multicolumn{1}{c@{}}{\textbf{RE}}
\\
\hline
\multicolumn{9}{@{}l@{}}{$n=200$}
\\
\quad$\hat{\mu}_{\mathrm{IPW}}$ & -0.74 & 12.62 & 12.64 & 5.06 & 28.65 & 179.02 &
181.30 & 53.01
\\
\quad$\hat{\mu}_{\mathrm{AIPW}}$ & 0.02 & 2.50 & 2.50 & 1.00 & -8.01 & 40.30 &
41.09 & 12.01
\\
\quad$\hat{\mu}_{\mathrm{OLS}}$ & 0.02 & 2.50 & 2.50 & 1.00 & -0.59 & 3.37 & 3.42 &
1.00
\\
\quad$\hat{\mu}_{\mathrm{IPW}-\mathrm{GBM}}$ & -3.37 & 3.11 & 4.59 & 1.86 & -4.36 & 3.13 &
5.37 & 1.57
\\
\quad$\hat{\mu}_{\mathrm{CAL},\mathrm{Q},\mathrm{DR}}$ & 0.02 & 2.50 & 2.50 & 1.00 & -2.13 & 3.26 &
3.89 & 1.14
\\
\quad$\hat{\mu}_{\mathrm{CAL},\mathrm{EL},\mathrm{DR}}$ & 0.02 & 2.50 & 2.50 & 1.00 & -2.73 & 3.98 &
4.83 & 1.41
\\
\quad$\hat{\mu}_{\mathrm{CAL},\mathrm{ET},\mathrm{DR}}$ & 0.02 & 2.50 & 2.50 & 1.00 & -2.40 & 3.48 &
4.23 & 1.24
\\
\quad$\hat{\mu}_{\mathrm{CAL},\mathrm{Q},\mathrm{MR}}$ & 0.02 & 2.50 & 2.50 & 1.00 & -1.23 & 2.84 &
3.09 & 0.90
\\
\quad$\hat{\mu}_{\mathrm{CAL},\mathrm{EL},\mathrm{MR}}$ & 0.02 & 2.50 & 2.50 & 1.00 & -1.13 & 3.00 &
3.20 & 0.93
\\
\quad$\hat{\mu}_{\mathrm{CAL},\mathrm{ET},\mathrm{MR}}$ & 0.02 & 2.50 & 2.50 & 1.00 & -1.17 & 2.86 &
3.09 & 0.90
\\[6pt]
\multicolumn{9}{@{}l@{}}{$n=1000$}
\\
\quad$\hat{\mu}_{\mathrm{IPW}}$ & 0.27 & 5.07 & 5.08 & 4.50 & 36.99 & 157.31 &
161.60 & 93.95
\\
\quad$\hat{\mu}_{\mathrm{AIPW}}$ & 0.01 & 1.13 & 1.13 & 1.00 & -13.38 & 72.19 &
73.42 & 42.69
\\
\quad$\hat{\mu}_{\mathrm{OLS}}$ & 0.01 & 1.13 & 1.13 & 1.00 & -0.86 & 1.49 & 1.72 &
1.00
\\
\quad$\hat{\mu}_{\mathrm{IPW}-\mathrm{GBM}}$ & -1.79 & 1.36 & 2.24 & 1.98 & -2.80 & 1.41 &
3.13 & 1.82
\\
\quad$\hat{\mu}_{\mathrm{CAL},\mathrm{Q},\mathrm{DR}}$ & 0.01 & 1.13 & 1.13 & 1.00 & -2.94 & 1.45 &
3.28 & 1.91
\\
\quad$\hat{\mu}_{\mathrm{CAL},\mathrm{EL},\mathrm{DR}}$ & 0.01 & 1.13 & 1.13 & 1.00 & -4.16 & 1.86 &
4.56 & 2.65
\\
\quad$\hat{\mu}_{\mathrm{CAL},\mathrm{ET},\mathrm{DR}}$ & 0.01 & 1.13 & 1.13 & 1.00 & -3.45 & 1.86 &
3.92 & 2.27
\\
\quad$\hat{\mu}_{\mathrm{CAL},\mathrm{Q},\mathrm{MR}}$ & 0.01 & 1.13 & 1.13 & 1.00 & -1.13 & 1.23 &
1.67 & 0.97
\\
\quad$\hat{\mu}_{\mathrm{CAL},\mathrm{EL},\mathrm{MR}}$ & 0.01 & 1.13 & 1.13 & 1.00 & -0.95 & 1.59 &
1.85 & 1.07
\\
\quad$\hat{\mu}_{\mathrm{CAL},\mathrm{ET},\mathrm{MR}}$ & 0.01 & 1.13 & 1.13 & 1.00 & -1.12 & 1.24 &
1.67 & 0.97
\\
\hline
\end{tabular*}
\end{table*}

We compared the performances of the inverse probability weighted
estimator $\hat{\mu}_{\mathrm{IPW}}$ and the\vadjust{\goodbreak} augmented inverse probability
weighted estimator
\begin{eqnarray*}
\hat{\mu}_{\mathrm{AIPW}}&=&\frac{1}{N}\sum_{i=1}^N
\frac{r_i}{\pi(x_i;\hat
{\beta})}y_i
\\
&&{}-\frac{1}{N}\sum
_{i=1}^N \biggl[\frac{r_i-\pi(x_i;\hat
{\beta})}{\pi(x_i;\hat{\beta})} \biggr]
\hat{m}(x_i),
\end{eqnarray*}
where $\hat{m}$ was the prediction from an ordinary least square
regression of $Y$ onto $Z$ for a correctly specified model and $X$ for
a misspecified model, the ordinary least square (OLS) estimator $\hat
{\mu}_{\mathrm{OLS}}=N^{-1}\times\sum_{i=1}^N \hat{m}(x_i)$ and the inverse
probability weighted estimator with a nonparametric propensity score
model fitted by generalized boosting machine (GBM) which was
implemented in the R package TWANG (McCaffrey, Ridgeway and
Morral,
\citeyear{mccaffrey2004propensity}). We used
GBM parameters suggested by Doctors Greg Ridgeway and Daniel McCaffrey
in a personal communication, with 3000 maximum iterations, a shrinkage
parameter of 0.005 and an iteration stopping rule that minimizes the
maximal marginal Kolmogorov--Smirnov\vadjust{\goodbreak} statistic. We denote the
corresponding inverse probability weighted estimates by $\hat{\mu
}_{\mathrm{IPW}-\mathrm{GBM}}$. We considered calibration estimators $\hat{\mu
}_{\mathrm{CAL},\mathrm{Q}}, \hat{\mu}_{\mathrm{CAL},\mathrm{EL}}, \hat{\mu}_{\mathrm{CAL},\mathrm{ET}}$ corresponding to
three special cases in the generalized empirical likelihood family:
Quadratic [Q: $\rho(v)=-(v+1)^2/2$], empirical likelihood [EL: $\rho
(v)=\ln(1-v)$] and exponential tilting [ET: $\rho(v)=-\exp(v)$]; we
also considered calibration estimators with one or two working outcome
regression models. With a single regression model, the calibration
estimators are doubly robust as an augmented inverse probability
weighted estimator. Multiple robust estimators calibrate to an
additional outcome model including all second and higher order
interactions of $Z$ for correctly specified models or $\sqrt{X}$ for
misspecified models. We chose the square-root transformation because
$X$ were positive and skewed to the right. We also considered the
logarithmic transformation and the results were similar. We used the
subscripts DR and MR to distinguish between the doubly robust and the
multiple robust calibration estimators.

Table~\ref{tab1} shows that both the augmented inverse probability weighted
estimator and the calibration estimators were more efficient than the
inverse probability weighted estimator. There are differences between
our results for the inverse probability weighted estimator and those in
\citet{kang2007demystifying}, which is due to the fact that the inverse
probability weighted estimator in our simulation is slightly different
from that discussed in \citet{kang2007demystifying}.
The inverse probability
weighted estimator considered in the simulations is shown in (\ref{E:ivee}). An
inverse probability weighted estimator considered by Kang and Schafer
replaced the denominator $N$ by $\sum_{i=1}^N r_i/\pi(x_i;\hat{\beta
})$. The two quantities should be close to each other when $N$ is large
and $\pi$ is correctly specified. In finite samples, however, the two
quantities can be quite different particularly when some $\pi(x_i)$
are close to zero. Both the \mbox{augmented} inverse probability weighted and
the calibration estimators had negligible biases and were efficient
when models were correctly specified. When models were \mbox{misspecified},
the augmented inverse probability weighted estimator had a considerable
bias and variability as shown in \citet{kang2007demystifying}, but the
calibration estimators, even the doubly robust ones, showed much better
performance compared to the augmented inverse probability weighted
estimator. The simulation scenario of \citet{kang2007demystifying} was
carefully designed such that the ordinary least squares estimator
outperforms all doubly robust estimators that were being considered.
The doubly robust calibration estimator, although substantially
improved over the augmented inverse probability weighted estimator, was
still inferior to the ordinary least squares estimator. Multiple robust
calibration estimators, however, outperformed the ordinary least
squares estimator in terms of mean squared error. This illustrates the
utility of multiple modeling. Although there is no guarantee that any
estimator dominates others when models are grossly misspecified, it is
likely that the true outcome model is better approximated by a
combination of multiple models rather than a single outcome model.
Within the generalized empirical likelihood family, choices of $\rho
(\cdot)$ did not affect the performance of the estimator in general.
An alternative way to improve the inverse probability weighted
estimator is to use a flexible nonparametric estimator of the
propensity score function, such as the generalized boosting machine
(McCaffrey, Ridgeway and Morral,
\citeyear{mccaffrey2004propensity}). However, inverse probability weighting with a
nonparametric method for propensity score estimation would induce more
small-sample bias than the parametric methods, and was less efficient
than calibration estimators in most cases.

\begin{table*}
\tablewidth=\textwidth
\tabcolsep=0pt
\caption{Comparisons among the calibration estimators and other
estimators under the Kang and Schafer scenario with interactions, \textup{(a)}
models in~Z, \textup{(b)} models in X. SSE represents the sampling standard
deviation, RMSE represents the root mean squared error and RE~represents~relative~efficiency which is the RMSE relative to $\hat{\mu
}_{\mathrm{OLS}}$}\label{tab2}
\begin{tabular*}{\textwidth}{@{\extracolsep{\fill}}ld{2.2}d{2.2}d{2.2}cd{3.2}d{3.2}d{3.2}d{2.2}@{}}
\hline
& \multicolumn{4}{c}{\textbf{(a)}} & \multicolumn{4}{c@{}}{\textbf{(b)}}
\\[-5pt]
& \multicolumn{4}{c}{\hrulefill} & \multicolumn{4}{c@{}}{\hrulefill}
\\
& \multicolumn{1}{c}{\textbf{Bias}} & \multicolumn{1}{c}{\textbf{SSE}}
& \multicolumn{1}{c}{\textbf{RMSE}} & \multicolumn{1}{c}{\textbf{RE}}
& \multicolumn{1}{c}{\textbf{Bias}} & \multicolumn{1}{c}{\textbf{SSE}}
& \multicolumn{1}{c}{\textbf{RMSE}} & \multicolumn{1}{c@{}}{\textbf{RE}}
\\
\hline
\multicolumn{9}{@{}l@{}}{$n=200$}
\\
\quad$\hat{\mu}_{\mathrm{IPW}}$ & -0.81 & 11.37 & 11.39 & 2.50 & 32.78 & 201.68 &
204.33 & 39.83
\\
\quad$\hat{\mu}_{\mathrm{AIPW}}$ & 0.25 & 4.56 & 4.57 & 1.00 & 6.12 & 80.46 & 80.63
& 15.72
\\
\quad$\hat{\mu}_{\mathrm{OLS}}$&3.17& 3.26&4.55&1.00&3.18&4.03&5.13&1.00
\\
\quad$\hat{\mu}_{\mathrm{IPW}-\mathrm{GBM}}$&-2.84&3.61&4.59&1.01&-3.36&3.69&4.99&0.97
\\
\quad$\hat{\mu}_{\mathrm{CAL},\mathrm{Q},\mathrm{DR}}$ & 0.51 & 3.45 & 3.49 & 0.77 & 0.36 & 4.08 &
4.10 & 0.80
\\
\quad$\hat{\mu}_{\mathrm{CAL},\mathrm{EL},\mathrm{DR}}$ & 0.42 & 3.56 & 3.58 & 0.79 & -0.21 & 4.15 &
4.16 & 0.81
\\
\quad$\hat{\mu}_{\mathrm{CAL},\mathrm{ET},\mathrm{DR}}$ & 0.47 & 3.48 & 3.51 & 0.77 & 0.10 & 4.10 &
4.11 & 0.80
\\
\quad$\hat{\mu}_{\mathrm{CAL},\mathrm{Q},\mathrm{MR}}$ & -0.05 & 2.74 & 2.74 & 0.60 & -0.24 & 3.31 &
3.32 & 0.65
\\
\quad$\hat{\mu}_{\mathrm{CAL},\mathrm{EL},\mathrm{MR}}$ & -0.05 & 2.74 & 2.74 & 0.60 & -0.23 & 3.45 &
3.45 & 0.67
\\
\quad$\hat{\mu}_{\mathrm{CAL},\mathrm{ET},\mathrm{MR}}$ & -0.05 & 2.74 & 2.74 & 0.60 & -0.22 & 3.34 &
3.35 & 0.65
\\[6pt]
%
%
%
\multicolumn{9}{@{}l@{}}{$n=1000$}
\\
\quad$\hat{\mu}_{\mathrm{IPW}}$ & 0.14 & 4.36 & 4.36 & 1.22 & 41.72 & 169.09 &
175.72 & 49.03
\\
\quad$\hat{\mu}_{\mathrm{AIPW}}$ & -0.09 & 2.74 & 2.74 & 0.77 & -11.97 & 44.03 &
45.63 & 12.30
\\
\quad$\hat{\mu}_{\mathrm{OLS}}$&3.20&1.55&3.56&1.00&3.03&1.89&3.58&1.00
\\
\quad$\hat{\mu}_{\mathrm{IPW}-\mathrm{GBM}}$&-1.45&1.54&2.12&0.60&-1.88&1.57&2.45&0.68
\\
\quad$\hat{\mu}_{\mathrm{CAL},\mathrm{Q},\mathrm{DR}}$ & 0.06 & 1.77 & 1.77 & 0.50 & -0.45 & 2.13 &
2.18 & 0.61
\\
\quad$\hat{\mu}_{\mathrm{CAL},\mathrm{EL},\mathrm{DR}}$ & 0.03 & 1.83 & 1.83 & 0.51 & -0.95 & 2.36 &
2.45 & 0.67
\\
\quad$\hat{\mu}_{\mathrm{CAL},\mathrm{ET},\mathrm{DR}}$ & 0.05 & 1.76 & 1.76 & 0.49 & -0.88 & 2.24 &
2.41 & 0.66
\\
\quad$\hat{\mu}_{\mathrm{CAL},\mathrm{Q},\mathrm{MR}}$ & {<}0.01 & 1.28 & 1.28 & 0.36 & 0.11 & 1.72 &
1.72 & 0.48
\\
\quad$\hat{\mu}_{\mathrm{CAL},\mathrm{EL},\mathrm{MR}}$ & {<}0.01 & 1.28 & 1.28 & 0.36 & 0.20 & 2.04
& 2.05 & 0.57
\\
\quad$\hat{\mu}_{\mathrm{CAL},\mathrm{ET},\mathrm{MR}}$ & {<}0.01 & 1.28 & 1.28 & 0.36 & 0.20 & 1.79
& 1.80 & 0.50
\\
\hline
\end{tabular*}      \vspace*{-3pt}
\end{table*}

Next, we performed additional simulations under a slight modification
of the Kang and Schafer scenario. The simulation setting was the same
as before except that an interaction term equal to $20Z_1Z_2$ was added
to the mean function of $Y$. We considered the same estimators as
discussed above. We presented the results in Table~\ref{tab2}. By comparing the
results of Tables~\ref{tab1} and~\ref{tab2}, we found that the performance of the
ordinary least squares estimator is sensitive to the specification of
the mean function, as illustrated in
Ridgeway and
\mbox{McCaffrey}
(\citeyear{ridgeway2007comment}). The
calibration estimator, on the other hand, still performed very well
under this modified scenario. In fact, the mean squared error of the
calibration estimators was substantially lower than other estimators.

In the rest of this section we focused on the Kang and Schafer scenario
without interaction. We examined the performance of the proposed
standard error estimator for the calibration estimators and the results
are shown in Table~\ref{tab3}, where the standard error estimates were close to
the sampling standard deviation and the empirical coverage of
approximate 95\% confidence intervals were close to their nominal levels.

\begin{table*}[b!]
\tablewidth=\textwidth
\tabcolsep=0pt
\caption{Performance of the standard error estimates of the
calibration estimators under the Kang and Schafer scenario: \textup{(a)} models
in Z, \textup{(b)} models in X. SSE represents the sampling standard deviation.
SEE represents the averaged standard error estimates. Coverage~(\%)~represents~the~empirical~coverage of approximate 95\% confidence
intervals}\label{tab3}
\begin{tabular*}{\textwidth}{@{\extracolsep{\fill}}lcccccc@{}}
\hline
& \multicolumn{3}{c}{\textbf{(a)}} & \multicolumn{3}{c@{}}{\textbf{(b)}}
\\[-5pt]
& \multicolumn{3}{c}{\hrulefill} & \multicolumn{3}{c@{}}{\hrulefill}
\\
& \textbf{SSE} & \textbf{SEE} & \textbf{Coverage (\%)}
& \textbf{SSE} & \textbf{SEE} & \textbf{Coverage (\%)}
\\
\hline
\multicolumn{7}{@{}l@{}}{$n=200$}
\\
\quad$\hat{\mu}_{\mathrm{CAL},\mathrm{Q}}$ & 2.50&2.56&96&3.04&2.95&95
\\
\quad$\hat{\mu}_{\mathrm{CAL},\mathrm{EL}}$ & 2.50&2.56&96&3.18&3.05&94
\\
\quad$\hat{\mu}_{\mathrm{CAL},\mathrm{ET}}$ & 2.50&2.56&96&3.09&2.95&94
\\[6pt]
\multicolumn{7}{@{}l@{}}{$n=1000$}
\\
\quad$\hat{\mu}_{\mathrm{CAL},\mathrm{Q}}$ & 1.13&1.15&96&1.29&1.30&91
\\
\quad$\hat{\mu}_{\mathrm{CAL},\mathrm{EL}}$ & 1.13&1.15&96&1.31&1.31&92
\\
\quad$\hat{\mu}_{\mathrm{CAL},\mathrm{ET}}$ & 1.13&1.15&96&1.29&1.30&92
\\
%
\hline
\end{tabular*}
\end{table*}

\begin{table*}
\tablewidth=\textwidth
\tabcolsep=0pt
\caption{Performance of the calibration estimators under correctly
specified or misspecified missing data models and multiple working
outcome regression models, \textup{(a)} one working model, \textup{(b)} two working
models, \textup{(c)} three working models and \textup{(d)} four working models.
SSE~represents~the~sampling standard deviation}\label{tab4}
\begin{tabular*}{\textwidth}{@{\extracolsep{4in minus 4in}}lcd{2.2}cd{2.2}cccd{2.2}c@{}}
\hline
& & \multicolumn{4}{c}{$\boldsymbol{n=200}$} & \multicolumn{4}{c@{}}{$\boldsymbol{n=1000}$}
\\[-5pt]
& & \multicolumn{4}{c}{\hrulefill} & \multicolumn{4}{c@{}}{\hrulefill}
\\
& & \multicolumn{2}{c}{\textbf{Correct}} & \multicolumn{2}{c}{\textbf{Misspecified}} &
\multicolumn{2}{c}{\textbf{Correct}} & \multicolumn{2}{c@{}}{\textbf{Misspecified}}
\\[-5pt]
& & \multicolumn{2}{c}{\hrulefill} & \multicolumn{2}{c}{\hrulefill} &
\multicolumn{2}{c}{\hrulefill} & \multicolumn{2}{c@{}}{\hrulefill}
\\
& & \multicolumn{1}{c}{\textbf{Bias}} & \multicolumn{1}{c}{\textbf{SSE}}
& \multicolumn{1}{c}{\textbf{Bias}} & \multicolumn{1}{c}{\textbf{SSE}}
& \multicolumn{1}{c}{\textbf{Bias}} & \multicolumn{1}{c}{\textbf{SSE}}
& \multicolumn{1}{c}{\textbf{Bias}} & \multicolumn{1}{c@{}}{\textbf{SSE}}
\\
\hline
$\hat{\mu}_{\mathrm{CAL},\mathrm{Q}}$ & (a) & 0.05 & 2.90 & -1.13 & 3.17 &  0.03 &
1.31 & -1.19 & 1.67
\\
& (b) & -0.10 & 2.79 & -2.18 & 3.03 &  0.02 & 1.26 & -2.26 & 1.53
\\
& (c) & 0.02 & 2.60 & -0.41 & 2.71 &  0.03 & 1.20 & -0.49 & 1.31
\\
& (d) & 0.02 & 2.50 & 0.02 & 2.50 &  0.01 & 1.13 & 0.01 & 1.13
\\[3pt]
$\hat{\mu}_{\mathrm{CAL},\mathrm{EL}}$ & (a) & 0.05 & 2.92 & -1.15 & 3.37 &  0.02 &
1.31 & -1.13 & 1.94
\\
& (b) & -0.10 & 2.80 & -2.24 & 2.91 &  0.02 & 1.26 & -2.27 & 1.76
\\
& (c) & 0.03 & 2.61 & -0.43 & 2.79 &  0.03 & 1.20 & -0.56 & 1.41
\\
& (d) & 0.02 & 2.50 & 0.01 & 2.49 &  0.01 & 1.13 & 0.01 & 1.13
\\[3pt]
$\hat{\mu}_{\mathrm{CAL},\mathrm{ET}}$ & (a) & 0.05 & 2.91 & -1.12 & 3.24 &  0.03 &
1.31 & -1.27 & 1.85
\\
& (b) & -0.10 & 2.79 & -2.18 & 3.07 &  0.02 & 1.26 & -2.24 & 1.65
\\
& (c) & 0.03 & 2.60 & -0.46 & 2.71 &  0.03 & 1.20 & -0.49 & 1.31
\\
& (d) & 0.02 & 2.50 & 0.02 & 2.50 &  0.01 & 1.13 & 0.01 & 1.13
\\
\hline
\end{tabular*}
\end{table*}

Next, we considered a case where the missing data mechanism was
possibly misspecified and multiple working outcome regression models
were assumed which may contain the correctly specified model. Let
$u_1=(1,Z_1)^T\hat{\gamma}_1$, $u_2=(1,Z_1,Z_2)^T\hat{\gamma}_2$,
$u_3=(1,Z_1,Z_2,Z_3)^T\hat{\gamma}_3$ and
$u_4=(1,Z_1,Z_2,Z_3, Z_4)^T\hat{\gamma}_4$, where $\hat{\gamma}_1$,
$\hat{\gamma}_2$, $\hat{\gamma}_3$ and $\hat{\gamma}_4$ were least
squares estimates obtained from complete case data. We considered
moment conditions from one to four working models: (a) one working
model $u=u_1$, (b) two working models $u=(u_1,u_2)$, (c) three working
models $u=(u_1,u_2,u_3)$ and (d) four working models
$u=(u_1,u_2,u_3,u_4)$. Only the fourth case contained the correctly
specified outcome regression model $u_4$. The simulation results are
shown in Table~\ref{tab4}. When multiple working outcome regression models were
assumed that contained the correct model, calibration estimators were
robust against misspecification of the missing data model and had
negligible bias. When missingness was correctly specified, inclusion of
more models decreased sampling variability. When missingness was
misspecified, the calibration estimators were slightly biased when
outcome models were misspecified, but sampling bias and variability
both decreased with an increasing number of models.

\begin{table*}[b]
\tablewidth=\textwidth
\tabcolsep=0pt
\caption{Performance of the estimators for two parameters, $\mu$ and
$p$. \textup{(a)} The inverse probability weighted estimator, \textup{(b)} the
calibration estimator using working model $m_1$, \textup{(c)} the calibration
estimator using working model $m_2$ and \textup{(d)} the calibration estimator
using working~models~$m_1$~and~$m_2$. SSE represents the sampling
standard deviation}\label{tab5}
\begin{tabular*}{\textwidth}{@{\extracolsep{\fill}}lcd{2.3}d{2.3}d{2.3}d{3.3}d{2.3}d{1.3}d{2.3}d{3.3}@{}}
\hline
& & \multicolumn{4}{c}{$\boldsymbol{n=200}$}
& \multicolumn{4}{c@{}}{$\boldsymbol{n=1000}$}
\\[-5pt]
& & \multicolumn{4}{c}{\hrulefill}
& \multicolumn{4}{c@{}}{\hrulefill}
\\
& & \multicolumn{2}{c}{\textbf{Correct}} & \multicolumn{2}{c}{\textbf{Misspecified}}
&  \multicolumn{2}{c}{\textbf{Correct}}
& \multicolumn{2}{c@{}}{\textbf{Misspecified}}
\\[-5pt]
& & \multicolumn{2}{c}{\hrulefill} & \multicolumn{2}{c}{\hrulefill}
&  \multicolumn{2}{c}{\hrulefill}
& \multicolumn{2}{c@{}}{\hrulefill}
\\
& & \multicolumn{1}{c}{\textbf{Bias}} & \multicolumn{1}{c}{\textbf{SSE}}
& \multicolumn{1}{c}{\textbf{Bias}} & \multicolumn{1}{c}{\textbf{SSE}}
& \multicolumn{1}{c}{\textbf{Bias}}
& \multicolumn{1}{c}{\textbf{SSE}} & \multicolumn{1}{c}{\textbf{Bias}}
& \multicolumn{1}{c@{}}{\textbf{SSE}}
\\
\hline
$\hat{\mu}$ & (a) & -0.74 & 12.62 & 28.65 & 179.02  & 0.27 & 5.07 &
36.99 & 157.31
\\
& (b) & 0.02 & 2.50 & 0.02 & 2.50  & 0.01 & 1.13 & 0.01 & 1.13
\\
& (c) & -0.27 & 3.18 & -1.36 & 3.69  & 0.01 & 1.54 & -0.47 & 2.86
\\
& (d) & -0.12 & 2.46 & -0.12 & 2.46  & 0.09 & 1.15 & 0.08 & 1.15
\\[3pt]
$\hat{p}$ & (a) & -0.003 & 0.064 & 0.104 & 0.616  & 0.001 & 0.027 &
0.129 & 0.515
\\
& (b) & -0.003 & 0.045 & 0.017 & 0.049  & {<}0.001 & 0.020 & 0.027 &
0.028
\\
& (c) & -0.001 & 0.034 & -0.002 & 0.034  & {<}0.001 & 0.013 & {<}0.001
& 0.015
\\
& (d) & -0.001 & 0.034 & -0.002 & 0.034  & {<}0.001 & 0.013 & {<}0.001
& 0.014
\\
\hline
\end{tabular*}
\end{table*}

\begin{table*}
\tablewidth=\textwidth
\tabcolsep=0pt
\caption{Analysis of the Washington basic health plan data. Relative
bias (RB) and relative efficiency (RE) of the following estimators: \textup{(a)}
the inverse probability weighted estimator, \textup{(b)} the calibration
estimator assuming a working linear model for conditional mean, \textup{(c)} the
calibration estimator assuming a working logistic model for conditional
proportion and \textup{(d)} the calibration estimator assuming both~working~models~for~conditional mean and conditional proportion}\label{tab6}
\begin{tabular*}{\textwidth}{@{\extracolsep{4in minus 4in}}ld{2.1}cd{2.1}cd{2.1}cd{3.1}c@{}}
\hline
&\multicolumn{4}{c}{$\boldsymbol{\hat{\mu}}$}&\multicolumn{4}{c@{}}{$\boldsymbol{\hat{p}}$}
\\[-5pt]
&\multicolumn{4}{c}{\hrulefill}&\multicolumn{4}{c@{}}{\hrulefill}
\\
&\multicolumn{2}{c}{\textbf{Correct}}&\multicolumn
{2}{c}{\textbf{Misspecified}}&\multicolumn{2}{c}{\textbf{Correct}}&\multicolumn
{2}{c@{}}{\textbf{Misspecified}}
\\[-5pt]
&\multicolumn{2}{c}{\hrulefill}
&\multicolumn{2}{c}{\hrulefill}
&\multicolumn{2}{c}{\hrulefill}&
\multicolumn{2}{c@{}}{\hrulefill}
\\
& \multicolumn{1}{c}{\textbf{RB (\%)}} & \multicolumn{1}{c}{\textbf{RE}} & \multicolumn
{1}{c}{\textbf{RB (\%)}} & \multicolumn{1}{c}{\textbf{RE}} & \multicolumn{1}{c}{\textbf{RB (\%)}}
& \multicolumn{1}{c}{\textbf{RE}} & \multicolumn{1}{c}{\textbf{RB (\%)}} & \multicolumn
{1}{c@{}}{\textbf{RE}}
\\
\hline
(a)& -0.3 & 1.00 & -9.1 & 1.00 & -0.2 & 1.00 &
-12.7 & 1.00
\\
(b)& {<}0.1 & 0.71 & {<}0.1 & 0.08 & -0.4 & 1.00 &
0.3 & 0.16
\\
(c)& -0.3 & 0.92 & -1.5 & 0.12 & -0.2 & 0.93 &
-0.7 & 0.15
\\
(d)&{<}0.1 & 0.70 & {<}0.1 & 0.08 & {<}0.1
& 0.93 &{<}0.1 &
0.15
\\
\hline
\end{tabular*}
\end{table*}

Next, we considered simultaneous estimation of two parameters of
interest, the sample mean $\mu$ and $p=P(Y>240)$. We assumed a working
model $m_1$ for $E(Y|Z)$ being a linear regression model with linear
predictors in $Z$ and a working model $m_2$ for $P(Y>240|Z)$ being a
logistic regression model with linear predictors in $Z$. Note that
$m_1$ is the true model for $E(Y|Z)$ but $m_2$ is not the true model
for $P(Y>240|Z)$. We considered the following four estimators: (a) the
inverse probability weighted estimator, (b) calibration estimator by
calibrating to predictions from $m_1$ only, (c) calibration estimator
by calibrating to predictions from $m_2$ only and (d) calibration
estimator by calibrating to predictions from both $m_1$ and $m_2$.
Since different choice of estimators within the generalized empirical
likelihood family gave similar results, we only reported the results
for $\rho(v)$ being a quadratic function. The simulation results are
given in Table~\ref{tab5}. When the missing data mechanism was correctly
specified, all estimators had small bias. When the missing data model
was misspecified, calibration estimators had much smaller biases
compared to the inverse probability weighted estimator. Similar to
Table~\ref{tab1}, calibration estimators had smaller sampling standard
deviations than the inverse probability weighted estimator. For the
estimation of $\mu$, efficiency of the calibration estimator was still
greatly improved compared to inverse probability weighted estimators
even when only a working model for $P(Y\geq240 | Z)$ was assumed.
However, the efficiency gain was less than the case when a working
model for $E(Y | Z)$ was assumed. Similar results held for the
estimation of $p$. When both models were assumed, the performance of
calibration estimators was no worse than the case when only one model
was assumed. By using a common set of weights calibrating to multiple
models, we achieved a similar improvement in efficiency relative to the
best improvement using different calibration weights for different estimands.

\subsection{Washington Basic Health Plan Data}

We performed an analysis using the Washington basic health plan data.
The data set contained information on a variety of health service
variables for 2687 households. For the purpose of illustration, we
chose an outcome $Y$ to be the total household expenditure on
outpatient visits, $X_1$ to be the family size and $X_2$ to be the
total number of outpatient visits. The distribution of medical
expenditure was highly skewed to the right with many zeroes. From the
full sample, the estimated mean household expenditure for outpatient
visits was $\mu_y=1948$ dollars, and the estimated proportion of
households with a total expenditure exceeding \$5000 was $p_y=0.1$. To
illustrate the performance of the calibration estimators, we compare
the results from the original data to simulated subsamples. Similar
analyses have been carried out in many survey sampling papers that
examined the performance of calibration estimators; see, for example,
Chen, Sitter and Wu
(\citeyear{chen2002using}) and Th\'{e}berge (\citeyear{theberge1999extensions}).
We drew a subsample following a
model $\operatorname{logit} P(R=1|X_1,X_2)=\beta_0+\beta_1X_1+\beta_2X_1I(X_1\geq
3)+\beta_3X_2$ and compared the performance of the inverse probability
weighted and the generalized empirical likelihood calibration
estimators for $\mu=E(Y)$ and $p=P(Y>5000)$ as if $Y$ were only
observed in the subsamples. The resampling process was repeated
$S=1000$ times.

We evaluated the estimators by comparing two performance measures,
percentage relative bias (RB\%) and relative efficiency (RE), defined by
\[
\mathrm{RB}_\sharp(\%)=\frac{1}{S}\sum
_{s=1}^S\frac{\hat{\mu}_{s,\sharp
}-\mu_y}{\mu_y}\times100
\]
and
\[
\mathrm{RE}_\sharp=\frac{\mathrm{MSE}_{\sharp}}{\mathrm{MSE}_{\mathrm{IPW}}},
\]
where $\hat{\mu}_{s,\sharp}$ is an estimator $\sharp$ (IPW or CAL)
computed from the $s${th} sample, $\mathrm{MSE}_{\sharp}=S^{-1}\sum_{i=1}^S(\hat{\mu}_{s,\sharp}-\mu_y)^2$ and $\mathrm{MSE}_{\mathrm{IPW}}$ is the MSE
of the corresponding inverse probability weighted estimators. The
performance of estimators were evaluated under both a correctly
specified missing data model and a misspecified working model $\operatorname{logit}
P(R=1|X_1,X_2)=\delta_0+\delta_1X_1+\delta_2X_1I(X_1\geq3)$. The
misspecified model ignored the dependence between the missingness
mechanism and $X_2$. For calibration estimators, we assumed a working
linear model for $E(Y|X_1,X_2)$ with a predictor linear in $X_1$ and
$X_2$, and a logistic regression model for $P(Y>5000|X_1, X_2)$ with a
predictor linear in $X_1$ and $X_2$. Note that both working models were
likely to be misspecified since the outcome data were not generated
from a known distribution. We considered calibration estimators using
only one working model assumption and using both model assumptions.
Since different choice of estimators within the generalized empirical
likelihood family gave similar results, we only reported the results
for $\rho(v)$ being a quadratic function. The results of the analyses
are shown in Table~\ref{tab6}.

When the missingness mechanism was correctly specified, all estimators
had a small bias, but the calibration estimators had improved
efficiencies relative to the inverse probability weighted estimators.
In the estimation of $\mu$, the efficiency of the calibration
estimator was still improved relative to the inverse probability
weighted estimators even when only a working model for $P(Y>5000|X)$
was assumed. However, the improvement in efficiency was less than the
case when a working model for $E(Y|X)$ was assumed. Similar results
held for the estimation of $p$. When both models were assumed, the
performance of the calibration estimator was no worse than the case
when only one model was assumed. This agrees with the theoretical
results in the paper. When the missing data mechanism was incorrectly
modeled, the inverse probability weighted estimator was severely biased
as expected, but all calibration estimators had small biases. This was
even true when the quantity being modeled was different from the
estimand. When both models were assumed, the performance of the
calibration estimator was no worse than the case when only one model
was assumed, and also had a negligible bias in the estimation of $\mu$
and $p$.

\section{Related Extensions}\label{sec8}

In this article we study the statistical properties of the generalized
empirical likelihood calibration estimators in the context of missing
data analysis. The calibration estimators allow multiple working
outcome
regression models to be assumed and enjoy an oracle property
where the same semiparametric efficiency bound is attained as if the
true outcome regression model is known in advance, when the missing
data mechanism is correctly specified. The estimators also enjoy a
multiple robustness property, where consistency holds when either the
missingness mechanism or any one of the working outcome regression
models is correctly specified. Calibration estimators provide an even
better protection against model misspecification than the existing
doubly robust estimators. Moreover, calibration allows the use of a
common set of weights in estimating multiple parameters and can improve
estimation efficiencies for multiple parameters simultaneously. In this
section we discuss several related extensions, including a different
but related way to construct calibration weights and an extension to
calibration estimating equations.

In previous sections we focus on a class of calibration estimators
satisfying moment conditions (\ref{E:cal}) which is related to many
existing estimators discussed in Section~\ref{sec6}. Other calibration
estimators can be constructed that satisfy (\ref{E:cal}) and enjoy
similar statistical properties as the proposed class. A different but
related calibration estimator can be constructed by noting that when
the missingness model is correctly specified we have
\[
E \biggl(\frac{R-\pi(X;\beta_0)}{\pi(X;\beta_0)}u(X) \biggr)=0.
\]
That is, $E(R\pi^{-1}(X;\beta_0)u(X)-u_\mu)=0$. We can define
calibration weights as
%
\begin{equation}
\label{E:cen1}\qquad p_i^*=\frac{1}{\pi(x_i;\hat{\beta})}\rho^{(1)} \bigl(
\hat{\lambda }_2^T \bigl(\pi^{-1}(x_i;
\hat{\beta})u(x_i)-\bar{u} \bigr) \bigr)
\end{equation}
for subjects with $r_i=1$, where
%
\begin{eqnarray}
\label{E:cen2}
 \hat{\lambda}_2&=&\arg\max_{\lambda}\sum
_{i=1}^N \rho \bigl(r_i
\lambda^T
\nonumber
\\[-8pt]
\\[-18pt]
&&\phantom{\arg\max_{\lambda}\sum
_{i=1}^N \rho \bigl(}{}\cdot\bigl(\pi^{-1}(x_i;\hat{
\beta})u(x_i)-\bar{u} \bigr) \bigr).\nonumber
\end{eqnarray}
In this case, we assume that $u$ contains a constant function. The
moment condition $\bar{u}=\sum_{i=1}^N r_ip_i^*u(x_i)$ is satisfied
from the first order condition of (\ref{E:cen2}). We can define a
calibration estimator to be $\hat{\mu}_{\mathrm{CAL}2}=\sum_{i=1}^N
r_ip_i^*y_i$. Suppose condition (\ref{E:drcond}) holds,
\begin{eqnarray*}
\hat{\mu}_{\mathrm{CAL}2}&=&\sum_{i=1}^N
r_ip_i^*y_i
\\
&=&\sum_{i=1}^N r_ip_i^*
\bigl(y_i-m_0(x_i)\bigr)+\sum
_{i=1}^N r_ip_i^*
m_0(x_i)
\\
&=&\sum_{i=1}^N r_ip_i^*
\bigl(y_i-m_0(x_i)\bigr)+\frac{1}{N}
\sum_{i=1}^N m_0(x_i),
\end{eqnarray*}
which converges in probability to $\mu$ by similar arguments as in
Section~\ref{sec4}. Therefore, the calibration estimator $\hat{\mu}_{\mathrm{CAL}2}$
enjoys a similar multiple robustness property enjoyed by the
calibration estimator $\hat{\mu}_{\mathrm{CAL}}$.



When we are interested in estimating a parameter $\theta_0$ defined\vspace*{1pt} by
an unbiased estimating function $g(y,x;\theta)$ such that
$E(g(Y,X;\theta_0))=0$, we can define $\hat{\theta}_{\mathrm{CAL}}$ to be the
solution of a calibration estimating equation ${g}_{\mathrm{CAL}}(\theta)=0$
where ${g}_{\mathrm{CAL}}(\theta)=\sum_{i=1}^N r_ip_ig(y_i,x_i;\theta)$.
Suppose $h_0(X)=E(g(Y,X;\theta_0)|X)$ exists and there exists
constants $a_0,\ldots,a_q$ such that $h_0(X)=a_0+\sum_{j=1}^qa_ju_j(X)$, then
\begin{eqnarray*}
{g}_{\mathrm{CAL}}(\theta)&=&\sum_{i=1}^N
r_ip_i\bigl(g(y_i,x_i;\theta
)-h_0(x_i)\bigr)
\\
&&{}+\sum_{i=1}^N
r_ip_i h_0(x_i)
\\
&=&\sum_{i=1}^N r_ip_i
\bigl(g(y_i,x_i;\theta)-h_0(x_i)
\bigr)
\\
&&{}+\frac{1}{N} \sum_{i=1}^N
h_0(x_i)
\end{eqnarray*}
and ${g}_{\mathrm{CAL}}(\theta_0)\convp0$ since $h_0(X)=E(g(Y,X;\theta_0)|X)$
and $E(h_0(X))=E(E(g(Y,X;\theta_0)|X))=0$. It follows from Newey and
McFadden (\citeyear{newey1986large}) that $\hat{\theta}_{\mathrm{CAL}}$ is a consistent estimate
of $\theta_0$ even when the missing data model is misspecified.

An associate editor suggested a possible alternative way of weighting
the individual working models and penalizing the misspecified models.
While this is an interesting idea, it is substantially different from
our methods. The calibration method put weights on individual
observations but not on models. This distinction is important in
Section~\ref{sec5} when we discuss multipurpose calibration. We showed that a
common set of weights can be used for efficient estimation of multiple
estimands. However, we believe that one cannot use a common set of
weights for penalizing individual models, because the correct models
are not the same for different estimands.

\section*{Acknowledgments}

This research was partially supported by the National Institutes of
Health grant.
The first author was supported in part by NIH Grants R01 AI089341 and
R01 DK07942. The second author was supported in part by The Hong Kong
RGC GRF 404012, The Chinese University of Hong Kong Direct Grant
2010/2011 Project ID: 2060422 and The Chinese University of Hong
Kong Direct Grant 2010/2011 Project ID: 2060444.
The authors would like to thank Professor Jon Wellner,
an associate editor and three reviewers for their insightful
comments and suggestions that greatly improved the paper.
The first author would like to thank Professor Norman
Breslow for helpful discussion and inspiration.
He also thanks Professor Mary Lou Thompson for helpful
comments and suggestions to improve the presentation of the paper.
The second author expresses his sincere gratitude to the hospitality
of both Hausdorff Center for Mathematics of the University of Bonn
and Mathematisches Forschungsinstitut Oberwolfach (MFO) in the
German Black Forest during the preparation of the present work.
The authors would like to thank Mr. Zheng Zhang for assistance.

\begin{supplement}[id=suppA]
\stitle{Proof of the Main Results\\}
\slink[doi]{10.1214/13-STS461SUPP} 
\sdatatype{.pdf}
\sfilename{sts461\_supp.pdf}
\sdescription{Online supplementary material is provided that
includes a list of regularity conditions, the proofs of Lemma~\ref{lem1},
Theorem~\ref{the2} and
Corollary~\ref{cor3}, together with two technical lemmas that
were needed to prove Lemma~\ref{lem1}.}
\end{supplement}

%

\end{document}